\documentclass[aps,prl,twocolumn,superscriptaddress,showpacs]{revtex4}
\usepackage{graphicx}
\preprint{Draft}
\newcommand{\beq}{\begin{eqnarray}}
\newcommand{\eeq}{\end{eqnarray}}
\newcommand{\vecdel}{{\boldmath $\delta$}}
\newcommand{\vecdeleq}{\mbox{\boldmath $\delta$}}
\newcommand{\smvecdeleq}{\mbox{\boldmath ${\scriptstyle \delta}$}}
\begin{document}

\title{Spin Waves in the Ferromagnetic Ground State\\
of the Kagom\'e Staircase System Co$_{3}$V$_{2}$O$_{8}$}

\author{M. Ramazanoglu}
\affiliation{Department of Physics and
Astronomy, McMaster University, Hamilton, Ontario, L8S 4M1, Canada}

\author{C.P. Adams}
\affiliation{Department of Physics and Astronomy, McMaster University,
Hamilton, Ontario, L8S 4M1, Canada}
\affiliation{Department of Physics, St.\ Francis Xavier University,
Antigonish, Nova Scotia, B2G 2W5 Canada}

\author{J.P. Clancy}
\affiliation{Department of Physics and
Astronomy, McMaster University, Hamilton, Ontario, L8S 4M1, Canada}

\author{A.J. Berlinsky}
\affiliation{Department of Physics and
Astronomy, McMaster University, Hamilton, Ontario, L8S 4M1, Canada}
\affiliation{Canadian Institute for Advanced Research, 180 Dundas St.\ W.,
Toronto, Ontario, M5G 1Z8, Canada}

\author{Z. Yamani} \affiliation{Canadian Neutron Beam Centre, National
Research Council, Chalk River Laboratories, Chalk River, Ontario, K0J
1P0, Canada}

\author{R. Szymczak}
\author{H. Szymczak}
\author{J. Fink-Finowicki}
\affiliation{Institute of Physics, Polish Academy of
Sciences, 02-668 Warsaw, Poland}

\author{B.D. Gaulin}
\affiliation{Department of Physics and Astronomy, McMaster University,
Hamilton, Ontario, L8S 4M1, Canada}
\affiliation{Canadian Institute for Advanced Research, 180 Dundas St.\ W.,
Toronto, Ontario, M5G 1Z8, Canada}
\date{\today}
\pacs{75.30.Ds, 75.50.Dd, 75.10.Dg}
\begin{abstract}
Inelastic neutron scattering measurements were performed on single
crystal Co$_{3}$V$_{2}$O$_{8}$ wherein magnetic cobalt ions reside on
distinct spine and cross-tie sites within kagom\'e staircase
planes. This system displays a rich magnetic phase diagram which
culminates in a ferromagnetic ground state below $T_{C} \sim 6$~K. We
have studied the low-lying magnetic excitations in this phase within
the kagom\'e plane.  Despite the complexity of the system at higher
temperatures, linear spin-wave theory describes most of the
quantitative detail of the inelastic neutron measurements. Our results
show two spin-wave branches, the higher energy of which displays
finite spin-wave lifetimes well below $T_C$, and negligible magnetic
exchange coupling between Co moments on the spine sites.
\end{abstract}
\maketitle

Magnetic materials in which the constituent magnetic moments reside on
networks of triangles and tetrahedra have been of great interest due
to their propensity for exotic ground states, a consequence of
geometrical frustration \cite{Diep}.
While ferromagnetically-coupled moments on
such lattices generally do not result in such ground states,
ferromagnets, and materials which display both ferromagnetic (FM) and
antiferromagnetic (AFM) interactions, on such lattices remain of great
interest, in part due to the relative scarcity of well-studied
examples, and in part due to intriguing spin ice \cite{Bramwell}
and multiferroic
phenomena \cite{multiferroic} which characterize some of these ground states.

The kagom\'e lattice is comprised of a two-dimensional network of
corner-sharing triangles.  Several realizations of magnetic moments on
stacked kagom\'e lattices with varying degrees of crystalline order have
been extensively studied.
Recently studied examples include jarosites, such as
KFe$_{3}$(OH)$_{6}$(SO$_{4}$)$_{2}$ \cite{Grohol} and
herbertsmithite ZnCu$_{3}$(OH)$_{6}$Cl$_{2}$ \cite{Helton},
both of which show evidence of strong magnetic frustration.

\begin{figure}
\includegraphics[scale=.6]{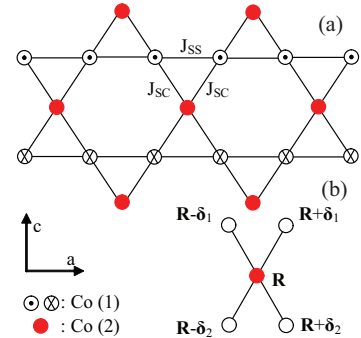}
\caption {\label{teori2} [color online]
(a) A schematic diagram of the kagom\'e staircase structure as
reduced to 2D in the $a$-$c$ plane.  The cobalt ions are
represented by open and solid circles for spine (M1) and cross-tie
sites (M2), respectively. Chains of spine sites
running parallel to the $a$-direction are alternatively
above ($\odot$) and below the plane ($\otimes$). (b) The basis used
in the linear spin-wave theory calculation.}
\end{figure}

The stacked kagom\'e staircase materials M$_{3}$V$_{2}$O$_{8}$
(M=Ni,Co) display orthorhombic crystal structures with space group
$C\, \! m \, \! c\, \! e$ \cite{Sauerbrei}.
Their kagom\'e layers are buckled and composed of edge-sharing
M$^{2+}$O$_{6}$ octahedra. These layers are separated by non-magnetic
V$^{5+}$O$_{4}$ tetrahedra. The buckled kagom\'e layers are
perpendicular to the orthorhombic $b$-axis and form what is known as a
stacked kagom\'e staircase structure.  Figure ~\ref{teori2} shows the
projection of the kagom\'e staircase onto the $a$-$c$ plane.  The two
inequivalent M sites are known as spines (M1) and cross-ties (M2).
The superexchange interaction between spine and cross-tie sites and
between two adjacent spine sites are denoted by $J_{sc}$ and $J_{ss}$,
respectively.

One member of this family, Ni$_{3}$V$_{2}$O$_{8}$ (NVO), undergoes a
series of phase transitions on lowering temperature
\cite{Kenzelmann,
Lawes, Wilson2, Rogado, Lancaster}.  A very interesting characteristic
of this compound is that it exhibits simultaneous ferroelectric and
incommensurate AFM order, that is, multiferroic behaviour, in one of
its ordered phases.  In isostructural Co$_{3}$V$_{2}$O$_{8}$ (CVO),
the $S=1$ magnetic moments at the Ni$^{2+}$ site are replaced
with $S=3/2$ Co$^{2+}$ moments. CVO also displays a rich
low temperature phase diagram, which has been studied using polarized
and unpolarized neutron diffraction, dielectric measurements
\cite{Chen}, magnetization and specific heat measurements
\cite{Szymczak}.  There is a series of four
AFM ordered phases below $T_N=11.3$~K
which can be characterized by incommensurate or commensurate ordering
wavevectors $(0 \tau 0)$.  In contrast to NVO,
the ultimate ground
state in CVO is ferromagnetic and the Curie temperature is
$T_C\sim6$~K.
Earlier powder neutron diffraction measurements \cite{Chen} on CVO
showed ordered magnetic moments of 2.73(2) and 1.54(6)~$\mu _B$
on the spine and cross-tie sites, respectively, at 3.1~K.
All moments are aligned along the
$a$-axis direction.
%

While much work has been performed on the phase diagrams
of NVO and CVO, little is known about the
excitations and, correspondingly, the
underlying microscopic spin Hamiltonian for these topical magnets.
The ferromagnetic state in CVO is an excellent venue for such a
study, as the ground state itself is very simple and therefore
the excitations out of the
ground state should be amenable to modeling.  In this Letter we
report inelastic neutron scattering (INS) measurements of the spin-wave
excitations within the kagom\'e staircase plane in the FM
ground state of single crystal CVO.  These measurements are compared
with linear spin wave theory which shows a surprising sublattice dependence to
the exchange interactions.

A large (5~g) and high-quality single crystal of CVO was grown using
an optical
floating-zone image furnace \cite{Szymczak}.  Thermal INS measurements
were performed at the Canadian Neutron
Beam Centre at the Chalk River Laboratories using the C5 triple-axis
spectrometer. A pyrolytic graphite (PG) vertically-focusing monochromator and
flat analyzer were used.  Measurements were performed with a fixed
final neutron energy of $E_f=13.7$~meV and a PG filter in the
scattered beam.  The collimation after the monochromator was
29'-34'-72' resulting in an energy resolution of 0.9~meV FWHM.

\begin{figure}
\includegraphics[scale=.38]{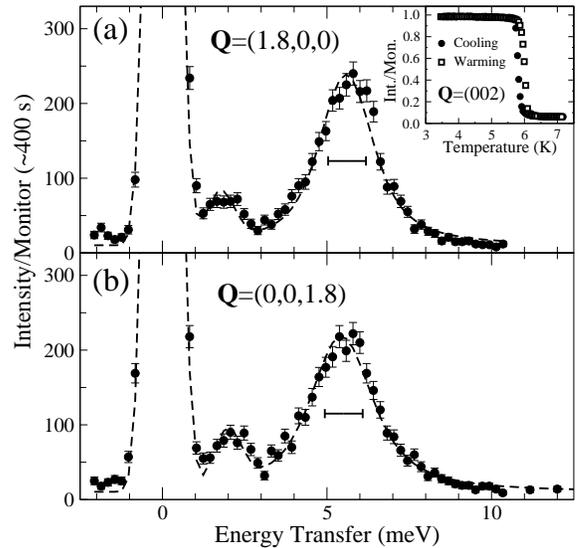}     
\caption{ \label{fig2} Representative constant-{\bf Q} INS spectra with
(a) {\bf Q}$=(1.8,0,0)$ and (b) {\bf Q}$=(0,0,1.8)$ at
$T=3$~K (FM phase). The broken line shows the resolution-corrected
fits to the data, as described in the text.  The solid horizontal bars
indicate the instrumental energy resolution.  The inset shows heating
and cooling scans of the (mainly) magnetic elastic (002) Bragg
scattering, characteristic of the FM ground state in CVO.}
\end{figure}

The crystal was oriented with the $(h 0 \ell)$ kagom\'e
staircase plane coincident with the scattering plane.  Constant-{\bf
Q} energy scans were performed along the high symmetry $(h00)$ and
$(0 0\ell)$ directions in this plane.  Figures 2 (a) and (b) show
representative
constant-{\bf Q} scans at $T=3$~K and {\bf Q}=(1.8, 0, 0) and (0, 0,
1.8), respectively.  The (002) nuclear Bragg peak is very weak,
and is coincident with a strong FM Bragg peak below
$T_C$.  The
inset of Fig.~\ref{fig2}(a) shows the temperature dependence of this
Bragg reflection on independent warming and cooling runs.  An abrupt
falloff in intensity near $T_C \sim 6$~K and accompanying hysteresis
indicate the strongly discontinuous nature of this phase transition.

The overall features of the two spectra in Fig. 2 are
quantitatively similar at $T=3$~K.  Two spin-wave
excitations, identified on the basis of their temperature and
$Q$-dependencies, are observed and have been modelled using
resolution-convoluted damped harmonic oscillator (DHO)
lineshapes \cite{shirane}.
The resulting fits are shown as the dashed lines in Fig. 2, and this
analysis allows us to conclude that the higher-energy spin wave, near
5.7~meV in both cases, has a substantial intrinsic energy width
of $\Gamma$=0.70(8)~meV at {\bf Q}=$(1.8,0,0)$ and $0.90(8)$~meV at
{\bf Q}=$(0,0,1.8)$.

A series of constant-{\bf Q} scans for $(h00)$ and $(00\ell)$
directions were collected at $T=3$~K
and
are presented as a color contour map in Figs. 3 (a) and (c).
Dispersive features corresponding to two bands of spin
waves are seen in both data sets. The top of the upper
spin-wave band at
$\Delta E\sim5.7$~meV corresponds to excitations
reported earlier using a time-of-flight technique \cite{Wilson2}.
These constant-{\bf Q} scans were fit to resolution-convoluted
DHO lineshapes, which gave intrinsic energy widths for the
higher-energy spin-wave mode at all wavevectors ranging from
$\Gamma = 0.6$ to 1.1~meV, while the lower-energy spin waves were
resolution limited at all wavevectors.  This indicates a finite
lifetime for the higher-energy spin waves even at temperatures
$\sim T_C/2$.

\begin{figure}
\includegraphics[scale=.43]{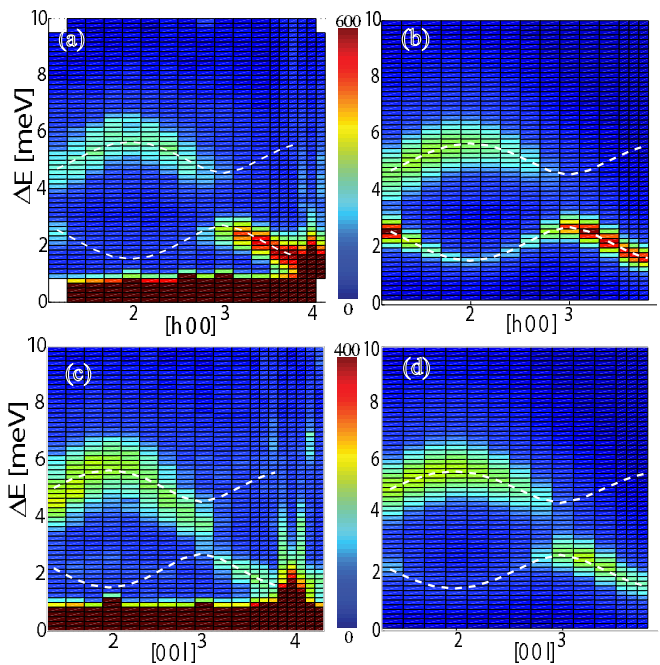}
\caption{ \label{teori1} Color contour maps of INS at $T=3$~K [(a) and
(c)] and corresponding linear spin-wave theory [(b) and (d)] as
described in the text.  The broken lines show the dispersion relations
resulting from this spin-wave theory analysis.}
\end{figure}

\begin{figure}
\includegraphics[scale=.38]{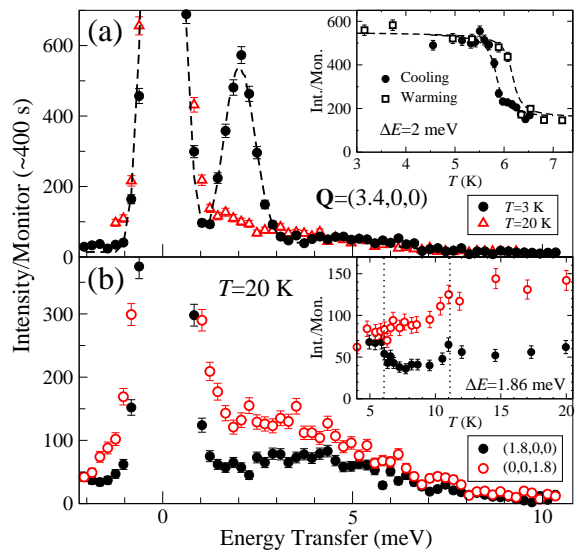}
\caption{ \label{fig3} [color online] (a) INS scans at
${\bf Q}=(3.4,0,0)$ for $T=3$~K and $T=20$~K. The inset shows
the temperature dependence of the scattering at the inelastic
peak position.
The lines are guides to the eye.
(b) Scans in the PM phase for ${\bf Q}=(1.8,0,0)$ and
${\bf Q}=(0,0,1.8)$.  The inset (same legend as main figure)
shows the temperature dependence
of the inelastic scattering at the spin-wave peak
(for $T=3$~K, see Fig.~\ref{fig2}) at $\Delta E=1.86$~meV for the different
$\bf{Q}$
directions.  The vertical dotted lines in the inset indicate $T_C$ and $T_N$. }
\end{figure}

The spin-wave spectrum evolves rapidly with increasing temperature.
Figure~\ref{fig3} (a) shows ${\bf Q}=(3.4,0,0)$ scans at $T=3$~K (FM
phase) and $T=20$~K [well into the paramagnetic (PM) phase].
In the FM phase there
is a prominent spin-wave peak at $\Delta E=2.0$~meV.  A higher energy
spin-wave peak is also present in this scan but with a much lower peak
intensity and an intrinsic energy width of $\Gamma=0.9$~meV.  At 20~K
the well-defined low-energy spin wave has disappeared and only broad
low-$\Delta E$ scattering remains.  The inset of Fig.~\ref{fig3} (a) shows
the temperature dependence of spin wave peak intensity at
$\Delta E=2.0$~meV in the neighborhood of $T_C$.  The same rapid falloff as
was seen in the magnetic Bragg scattering at $(002)$ is seen in the
inelastic intensity, as well as the same hysteresis, indicating that these
spin waves are strongly coupled to the ferromagnetic order parameter.

Figure~\ref{fig3} (b) shows the same constant-{\bf Q} scans as in
Fig.~\ref{fig2} but now at 20 K rather than 3 K.  The well-defined
spin-wave modes are no longer present, and the low-energy inelastic
scattering is significantly greater at ${\bf Q}$=$(0,0,1.8)$ as
compared with $(1.8,0,0)$ at $T=20$~K.  The temperature dependence
of this scattering is
contrasted in the inset of Fig.~\ref{fig3} (b). The {\bf
Q}=$(1.8,0,0)$ inelastic scattering drops quickly at $T_C\sim6$~K
while that at {\bf Q}= $(0,0,1.8)$ gradually increases with
temperature.  We attribute this difference to prominent longitudinal
easy-axis spin fluctuations along the $a$-axis at high temperatures.
Given that the neutron scattering cross-section \cite{shirane} is sensitive to
magnetic fluctuations perpendicular to {\bf Q}, longitudinal
fluctuations would be seen
in the {\bf Q}=$(0,0,1.8)$ spectrum rather than the $(1.8, 0, 0)$ spectrum.
Note that the highest temperature
transition, to the paramagnetic state at $T_N=11.3$~K, is evident as a
change in slope of the temperature dependence
shown in the inset of Fig.~\ref{fig3} (b), and both $T_C \sim 6$~K
and $T_N=11.3$~K are indicated by dashed vertical lines in this inset.

We have carried out a linear spin-wave theory analysis of the magnetic
excitations observed in Figs. 3 (a) and (c) to understand as much of the
relevant microscopic spin Hamiltonian as possible.  The full
Hamiltonian is potentially complicated if account is taken of the two
inequivalent magnetic sites and the 3D kagom\'e staircase
structure. We employed a 2D model in which the magnetic ions in a
layer are projected onto the average plane of the layer
(Fig.~\ref{teori2}) and
only near-neighbor exchange and single-ion anisotropy are included.

We change from the conventional centered-rectangular unit cell to
a primitive rhombohedral unit cell defined by lattice
vectors
${\bf R}_1=(\frac{a}{2}, \frac{c}{2})$ and
${\bf R}_2=(\frac{a}{2},-\frac{c}{2})$ and basis vectors
\vecdel$_1={\bf R}_1/2$,
\vecdel$_2={\bf R}_2/2$, and \vecdel$_3=\mbox{\boldmath $0$}$.
If ${\bf R}$ describes the
set of lattice points and ${\bf S}_{\bf r}$ is the
spin operator at a location {\bf r} we can write
\beq
\label{hamil1}
{\cal H}_{sc} & = & -J_{sc} \sum_{\bf R} \sum_{\smvecdeleq = \pm
  \smvecdeleq_1,\pm \smvecdeleq_2} {\bf S}_{\bf R} \cdot {\bf S}_{{\bf
  R}+{\smvecdeleq}},\\
{\cal H}_{ss} & = & -J_{ss} \sum_{\bf R} \big[{\bf
  S}_{{\bf R}+ \smvecdeleq_1} \! \cdot \! {\bf S}_{{\bf R}-\smvecdeleq_2} +
  {\bf S}_{{\bf R}- \smvecdeleq_1} \! \cdot \! {\bf S}_{{\bf
  R}+\smvecdeleq_2} \big].
\eeq
${\cal H}_{sc}$ and ${\cal H}_{ss}$
are the exchange interactions between spine and cross-tie spins with
couplings $J_{sc}$ and $J_{ss}$.  The fact that the spine and
cross-tie spins are found to be ferromagnetically aligned implies
that $J_{sc}$ is positive. We choose the spin $z$-axis parallel to
the crystallographic $a$-axis, consistent with the ordered moment
direction.  Magnetization measurements have established that the
crystallographic $b$-axis (spin $x$-axis) is a harder axis than the
$c$-axis (spin $y$-axis) \cite{Szymczak}.  The appropriate
single-ion anisotropy Hamiltonian, ${\cal H}_a$, distinguishes the
three orthogonal directions and the inequivalent sites \beq
\label{hamilc}
{\cal H}_a = \sum_{\bf R} \sum_{i=1,2,3} \sum_{\alpha = x,y,z}
 A_{i}^{\alpha}( S^{\alpha}_{{\bf R}+\smvecdeleq _i})^2.
\eeq
The sequence from hard to
easy axis is established by the condition
$A_{i}^{x} > A_{i}^{y} > A_{i}^{z}$.
We use reduced
anisotropies: $\Delta_i = (A_i^x + A_i^y -2A_i^z)/2$ and
$\epsilon_i = (A_i^x - A_i^y)/2$.  Since the $i=1$ and $i=2$
sites are equivalent there are 4 independent parameters
$\Delta_s$, $\Delta_c$, $\epsilon_s$, and $\epsilon_c$.

In linear spin-wave theory \cite{holstein}
the total Hamiltonian ${\cal H}$ may be
written in term of spin-wave operators
$c_{i}({\bf k})$ and
$c^{\dagger}_{i}({\bf k})$
\begin{widetext}
\beq
\label{quadH}
{\cal H} = \sum_{{\bf k},i,j} h_{ij}({\bf k}) c^{\dagger}_i ({\bf k})
  c_{j}({\bf k})\
   + \frac{1}{2}\sum_{{\bf k},i,j} g_{ij}({\bf k})
  \left(c^{\dagger}_i({\bf k}) c^{\dagger}_j(-{\bf k})
       +c_j(-{\bf k}) c_i({\bf k})\right),
\eeq
\beq
h({\bf k})= 2S
\left(
\begin{array}{ccc}
J_{sc}+J_{ss}+\Delta_s &
  -J_{ss} \cos {\bf k}\cdot(\vecdeleq_1+\vecdeleq_2) &
  -J_{sc} \cos{\bf k} \cdot \vecdeleq_1 \\
-J_{ss}\cos{\bf k} \cdot (\vecdeleq_1 +\vecdeleq_2) &
   J_{sc}+J_{ss}+\Delta_s &
  -J_{sc} \cos{\bf k}\cdot \vecdeleq_2 \\
-J_{sc} \cos{\bf k} \cdot \vecdeleq_1 &
  -J_{sc} \cos{\bf k}\cdot \vecdeleq_2 &
  2J_{sc}+\Delta_c
\end{array}
\right), \qquad
g({\bf k})=2S\left(\begin{array}{ccc}
\epsilon_s & 0 & 0 \\
0 & \epsilon_s & 0 \\
0 & 0 & \epsilon_c \\
\end{array}\right),
\eeq
\end{widetext}
where $i,j=1,2,3$.
Although the moments on the different sites are unequal
we make the simplifying assumption that $S=3/2$ for all sites.

The unpolarized magnetic neutron scattering cross-section
and the spin-spin correlation functions it contains can be related to
one spin-wave Green's functions.  The Green's functions can be
calculated by inverting a matrix involving the two matrices, $h({\bf
k})$ and $g({\bf k})$ defined above, following Coll and Harris
\cite{coll}.  The spin-wave energies are determined by solving for the
zeros of a determinant involving matrix elements of the inverse
Green's function.  The resulting
scattering function ${\cal S}({\bf Q},\Delta E)$
for {\bf Q} parallel to $(00 \ell)$ is
proportional to \beq
\label{intensity}
{\cal S}({\bf Q},\Delta E) \! \propto \!
 \sum_{i,j}  \mbox{Im}  \! \left[ (h \!- \!g)
(z^2 I \! - \! h^2 \! + \! g^2 \! + \! hg \! - \! gh)^{-1} \right]_{i,j}
\eeq
where $I$ is the identity matrix, $\bf{k}=\bf{Q}$
and $z=\Delta E-i \Gamma$ is the complex
energy, and $i$ and $j$ are summed over the indices of the $3 \times 3$
matrix in square brackets.  The expression for {\bf Q} parallel to
$(h00)$ has a similar form.
The resulting scattering functions are multiplied by the magnetic form factor,
the Bose factor, and a single intensity scale factor, and are plotted in Fig.~\ref{teori1} (b) and (d).

Best agreement between the experimental data and the
spin-wave theory calculation in Fig.~\ref{teori1} was obtained for
magnetic coupling predominantly between the spine and cross-tie Co
ions with $J_{sc}=1.25\pm0.08$~meV, while the spine-spine coupling
$J_{ss}$ vanishes.  The best fit spin-wave uniaxial anisotropy
parameters are $\Delta_{s}=1.5\pm0.1$~meV,
$\Delta_{c}=2.13\pm0.2$~meV, $\epsilon_{s}=0.6\pm0.3$~meV, and
$\epsilon_{c}=1.2\pm0.3$~meV.

Figure~\ref{teori1} shows that the spin-wave theory gives a very good
description of the dispersion of the two modes (dashed lines) and
accounts for the observed trend of the spin waves to trade intensity
as a function of $\bf{Q}$.
This description is not perfect, however.  The calculated dispersion
of the lower spin-wave band is low compared with
experiment near (200) and (002) where the intensity is very weak.
The calculation is not convolved with the instrumental resolution;
instead the energy width is manually set in both high and
low-energy bands to correspond to the
measured width.  The broad (in energy)
neutron groups corresponding to the upper spin-wave bands are most
evident near $(200)$ and $(002)$.  The lower energy spin-wave band
becomes much more intense near the zone centers of $(400)$ and
$(004)$. Steep excitation branches near (400) and (004) with
comparitively weak intensity in the
experimental data [Figs. 3 (a) and (c)] are identified as
acoustic phonons with a speed of sound of $1050 \pm 100$~m/s, in
both directions.


To conclude, our INS study of the spin-wave excitations in the
ferromagnetic ground state of CVO within its kagom\'e staircase plane
reveal two separate spin-wave bands between 1.6 and 5.7~meV.  The
upper spin-wave band is damped with finite energy widths $\Gamma$ in
the range of 0.6 to 1.1~meV.  These spin-wave excitations can be
accurately described by a simple model Hamiltonian and linear
spin-wave theory.  The model gives a magnetic coupling that is
predominantly between the spine and cross-tie sites of the kagom\'e
staircase.

\end{document}